\documentclass[%
twocolumn,showpacs
 amsmath,amssymb,
prb,
]{revtex4-1}

\usepackage[utf8]{inputenc}
\usepackage{graphicx}
\usepackage{dcolumn}
\usepackage{bm}

\usepackage{subcaption}

\begin{document}

\title{Density-functional investigation of molecular graphene: CO on Cu(111)}

\author{Matti Ropo$^{1,2}$}
\email{matti.ropo@tut.fi}
\author{Sami Paavilainen$^1$}
\author{Jaakko Akola$^{1,2}$}
\author{Esa Räsänen$^1$}
\affiliation{%
 $^1$ Department of Physics, Tampere University of Technology, P.O. Box 692, FI-33101 Tampere, Finland\\
 $^2$ COMP Centre of Excellence, Department of Applied Physics, Aalto University, FI-00076 Aalto, Finland
}%

\date{\today}

\begin{abstract}
Man-made artificial graphene has attracted significant attention in the past few years 
due to the possibilities to construct designer Dirac fermions with unexpected topological
properties and applications in nanoelectronics. Here we use a first-principles
approach within density-functional theory to study molecular graphene similar to the 
experiment by Gomes~{\em et al.}, Nature {\bf 483}, 306 (2012). The system comprises carbon 
monoxide molecules arranged on a copper (111) surface in such a way that a hexagonal lattice 
is obtained with the characteristic electronic properties of graphene. Our results show in detail how 
carbon monoxide molecules modify the copper surface (and regions beneath) and create a hexagonal 
lattice of accumulated electrons between the adsorbate molecules. We also demonstrate how the 
properties of the formed Dirac fermions change as the CO density is tuned, and provide a 
direct comparison with experimental scanning tunneling microscope images.
\end{abstract}

\maketitle


Graphene~\cite{graphene} has raised enormous interest in the scientific community during the last decade.  
In many ways, this is due to its peculiar electronic structure which arises from the honeycomb lattice. 
Recently, research groups have shown analogous properties on different systems~\cite{nature497_422} with sixfold 
symmetry ranging from molecules on metal surfaces~\cite{nature483_306} to trapped atoms,~\cite{tarruell}
and further to semiconductor heterostructures~\cite{gibertini,singha,esa} and nanocrystals.~\cite{kalesaki}
These realizations of ``artificial graphene''~\cite{polini} make it possible to investigate and exploit 
the unique characteristics of graphene with tunable parameters which are not accessible or 
very difficult to change with real graphene.

Recently, Gomes {\em et al.}~\cite{nature483_306} constructed molecular graphene by placing 
carbon monoxide (CO) molecules on copper (111) surface in a controlled manner.  The adsorbate 
molecules create artificial constraints for the delocalized surface electron density and transforms 
the two-dimensional (2D) electron system of the triangular Cu(111) lattice to a hexagonal one.  This 
(2D) honeycomb lattice is tunable by changing the density and placing of the CO molecules. In fact,
Gomes {\em et al.}~\cite{nature483_306} not only found signatures of Dirac physics in the surface-state
conductance, but they were also able to manipulate the band structure by creating Kekul\'e distortion
as well as a pseudomagnetic field through strain. At present, of all the possibilities available
for the generation of artificial graphene (see above), the molecular approach appears
as the most promising in terms of controllability and applications.

In addition to molecular graphene, there has been substantial interest in carbon monoxide on metal surfaces 
both experimentally and theoretically,~\cite{jacs107_578,ss59_593,prl76_2141,prb76_195440,ss590_117} 
and the interaction between the molecule and metal surfaces is well understood.~\cite{jacs107_578,prl76_2141} 
The chemical bonding can be mainly ascribed to the interaction 
between the 5$\sigma$ and 2$\pi^{*}$ states of CO and the $d$-states of the copper surface, while 
there is no significant charge transfer. Despite the well-understood nature of this interaction, we are 
not aware of any computational investigations of this system in the first-principles level in 
order to study how electronic properties of graphene arise for CO covered Cu(111) 
surfaces.

Here, we apply the density-functional theory (DFT) to study CO on Cu(111) surfaces with similar setups 
as in the experiment by Gomes \emph{et al.}~\cite{nature483_306} In particular, we investigate at the atomistic 
level how CO modifies the electronic states of the copper surface. Therefore, in terms of modeling the 
experiment, our first-principles approach goes significantly beyond the 2D model applied for finite flakes of 
molecular graphene in Ref.~\onlinecite{flake}. We are able to reproduce the signatures of Dirac fermions in 
the system and find that the adsorbate density affects the location of the Dirac crossing with respect to the 
Fermi level, as the surface electrons are pushed to a hexagonal lattice constrained by CO molecules. To 
analyze the structure furher, we examine the electron localization function (ELF) as a measure for electron 
pairing and show that the surface electrons indeed have a delocalized, metallic character.

Our DFT simulations are performed using the FHIaims~\cite{cpc180_2175} program with the numerical 
atomic orbital Tier 2 level basis set. For the exchange-correlation functional we use the 
Perdew–Burke–Ernzerhof (PBE) form of the generalized-gradient approximation.~\cite{prl77_3865}
The Cu(111) surface is modeled using a slab geometry with nine atomic layers of Cu, and the periodically 
repeated slabs are separated by a 22\,Å thick vacuum. The lateral sizes of the Cu slab systems are coupled 
to the CO coverage, thus resulting in simulation boxes with different sizes and numbers of atoms
for the $1\times1$,$\sqrt{3}\times\sqrt{3}$, $2\times2$, $3\times3$ and $4\times4$ system studied.
For each CO coverage, the CO molecule is positioned on the top site of one surface Cu atom 
of the corresponding unit cell (three on the $\sqrt{3}\times\sqrt{3}$ surface). The atoms in the 
lowermost four atomic layers opposite to the surface containing CO molecules have been fixed to 
the positions bulk copper. All other atomic positions have been optimized during the simulations.

The ELF analysis is performed within the VASP program.~\cite{prb54_11169} Here the 2$\times$2 
geometry obtained from the FHIaims calculations is used as input, and the calculations 
are carried out with 157 irreducible {\it k}-points. We use plane waves with a cut-off 
energy of 500~eV and the projector augmented wave (PAW) method~\cite{prb59_1758,prb50_17953} 
with the PBE. The band structure obtained from VASP is nearly identical with that of FHIaims, and VASP 
was used for further analysis. 



The number of {\bf k}-points in the $xy$-plane, molecular density and CO-CO molecule distance for 
each surface coverage are given in Table \ref{tab:numprop}. In order to 
avoid artificial strain, a theoretical Cu lattice parameter of 3.63\,Å is used which 
is 0.5\% larger than the experimental value of 3.61\,Å (Ref.~\onlinecite{aca25_676}). The covalent C-O distance 
of the adsorbed CO molecule is 1.15\,Å in each case and the C-Cu distance (adsorbate-substrate) 
is 1.85\,Å for all surfaces except for $1\times1$ where it is slightly larger (1.88\,Å). Previously, 
Gajdos and Hafner~\cite{ss590_117} have reported 1.156\,Å for C-O and 1.86\,Å for C-Cu distance 
with the PW91 functional, whereas Stroppe \emph{et al.} \cite{prb76_195440} obtained 1.158\,Å 
for the C-O distance with the PBE functional.  Both of these studies were performed for a $2\times4$ 
surface coverage. Due to the interaction with CO, the copper atom directly below the molecule is 
elevated between 0.16\,Å  and 0.12\,Å (from $\sqrt{3}\times\sqrt{3}$ to $4\times4$ coverage) 
higher than rest of the surface atoms, which is similar to a height difference of  0.124\,Å  between 
the outermost and innermost metal atoms in the first layer reported by Stroppe {\em et al.}~\cite{prb76_195440}

\begin{table*}
\begin{tabular}{|c|c|c|c|c|c|c|}
\hline
System & \#k-points & CO dens. (mol./Å$^2$) & CO-CO dist.(Å) & C charge (e) & O charge (e) 
&  $E_{\textrm{Ads.}}$ (meV) \\
\hline
$1\times1$ & 365 & 0.175 & 2.57 & 0.018 &  -0.071 & 186 \\
$\sqrt{3}\times\sqrt{3}$ & 221  & 0.058 & 4.49 & 0.042 &  -0.083  & -765 \\
$2\times2$ & 365 & 0.044 & 5.14 & 0.045 & -0.083 & -762\\
$3\times3$ & 221 & 0.019 & 7.70 & 0.051 & -0.078 & -742\\
$4\times4$ &  61 & 0.011 & 10.27 & 0.049 & -0.080 & -747\\
\hline
\end{tabular}
\caption{Number of {\bf k}-points, molecular surface density, CO-CO distance, Hirshfeld 
charges\cite{tca44_129} for C and O, and absorption energies for investigated surface coverages.
\label{tab:numprop}}
\end{table*}


The adsorption energies of CO molecules on Cu(111) surface are calculated using following 
equation:
\begin{equation}
E_{\rm Ads.} = E_{\rm CO+Cu} - E_{\rm Cu} - E_{\rm CO}.
\end{equation}
where $E_{\rm Ads.}$, $E_{\rm CO+CU}$, $E_{\rm Cu}$, and $E_{\rm CO}$ are the adsorption energies of a CO 
molecule on a top site of Cu(111) surface, energy of the Cu slab with CO, without CO, and the 
energy of a free CO molecule, respectively. The adsorption energies are shown in Table~\ref{tab:numprop}, 
and for the most surface coverages they are around 
-750\,meV, but for the $1\times1$ coverage the energy is positive indicating that it is not a stable
configuration due to the CO-CO repulsion. Previously, Stroppa \emph{et al.}\cite{prb76_195440} 
reported an adsorption energy of -709\,meV with the PBE functional. 
The discrepancy between $1\times1$ and other coverages, which was already noticed in the C-Cu 
distances, is also seen in the charge transfer of CO (Table~\ref{tab:numprop}): C atoms in the 
$1\times1$ surface have an effective charge of only 0.018\,e, whereas the other systems show
values of $\sim$0.05\,e.

\begin{figure}
    	\includegraphics[scale=0.25]{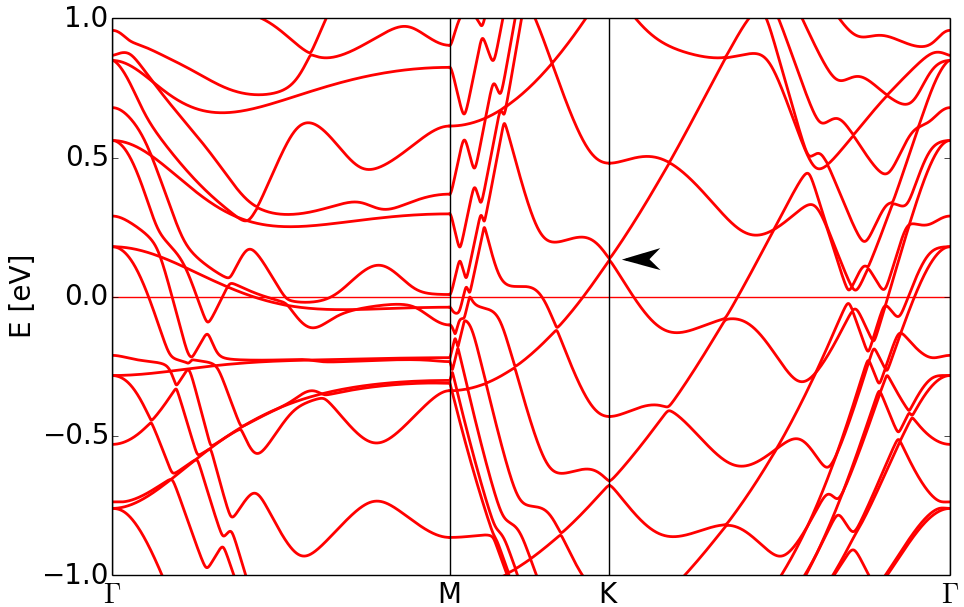}
        \caption{Band structure of the $2\times2$ surface system. Two band crossings at the $K$ 
        point are clearly visible, one just above the Fermi level and the second one further in the 
        valence band. The closest band crossing to the Fermi level can be found at 130\,meV indicated
by an arrow.
        \label{fig:band_2}}
\end{figure}
\begin{figure}
    	\includegraphics[scale=0.25]{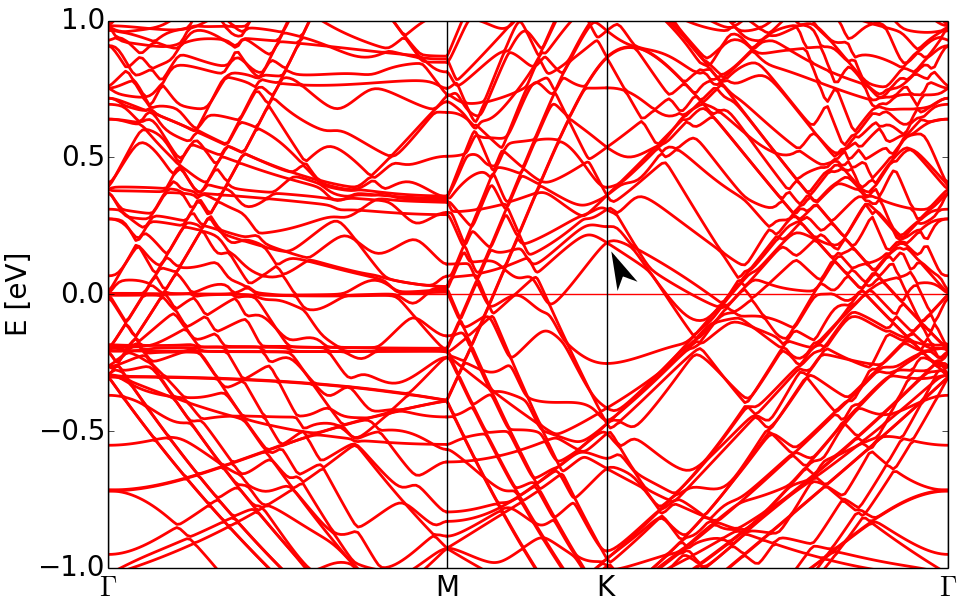}
        \caption{Band structure of the $4\times4$ surface system. The crossing closest to the Fermi 
        level is located at 190\,meV (arrow). In this case, the band structure is more difficult to interpret than in the 
$2\times2$ case as the system size (number of Cu atoms in the simulation box) is significantly larger, leading to more
bands.
        \label{fig:band_4c}}
\end{figure}

The band structures of the investigated systems reveal a band crossing that resembles the famous 
Dirac cone located at the $K$ symmetry point of the Brillouin zone. The position (energy) of the
closest crossing with respect to the Fermi level depends on the CO coverage, and the exact 
positions are given in Table~\ref{tab:fermivelo}. The band structures of the $2\times2$ and $4\times4$ 
systems are shown in Figs.~\ref{fig:band_2} and \ref{fig:band_4c}. For the $1\times1$ coverage, we 
observe a clear Dirac cone 0.92 eV above the Fermi energy. In all other cases as the CO coverage 
reduces, we observe multiple band crossings due to the laterally increasing system size which 
comprises multiple Cu unit cells. Correspondingly, the band structure becomes more difficult to 
analyze as the the number of electrons (bands) increases (cf. Figs.~\ref{fig:band_2} and 
\ref{fig:band_4c}). The number of bands is the largest for the 4$\times$4 coverage, where the 
CO-CO distances are the largest.

We have performed further band-structure analysis for the 2$\times$2 surface with the VASP 
program in order to study the nature of the Kohn-Sham (KS) states close to band-crossing point 
above the Fermi level. The real space projections of the KS states as a function of wave vector 
{\bf k} verify that the symmetry of the KS states changes at the K-point proving that the 
band crossing is genuine. The states corresponding to the band-crossing point have contribution 
from both CO and Cu orbitals, but they are not solely localized to the (111) surface. Instead, they have 
significant weight also deeper in the copper slab. The projected density of states onto atomic 
orbitals reveals that the CO contribution is still much smaller than Cu contribution for the crossing 
bands, i.e. the Dirac cone is mostly associated with copper. On the contrary, the contribution of
CO increases for the KS states higher in the conduction band, and the next band crossing 
at 1.12\,eV is associated with CO.

We can estimate the Fermi velocities of electrons from the slopes of the bands at the crossing points. 
The velocities range between $3.0 - 6.0\times 10^5$ m/. As an important factor in the validation of the 
present model and simulations, our range is of the same order as the value $6.45\times10^5$ m/s estimated 
by Gomes \emph{et al.}\cite{nature483_306} in the experiments for the same system.

\begin{table}
\begin{tabular}{|c|c|c|}
\hline
System  &Fermi velocity(m/s) & DP pos. (meV)\\
\hline
$1\times1$ & $6.0\times10^{5}$ & 920 \\
$\sqrt{3}\times\sqrt{3}$ & $5.3\times10^{5}$ & -110 \\
$2\times2$ & $3.5\times10^{5}$ & 130 \\ 
$3\times3$& $5.2\times10^{5}$ & -50 \\
$4\times4$ & $3.0\times10^{5}$ & 190 \\
\hline
\end{tabular}
\caption{Fermi velocity of electrons and the position of the Dirac point for different CO 
coverages.
\label{tab:fermivelo}}
\end{table}

\begin{figure}
\includegraphics[scale=0.22]{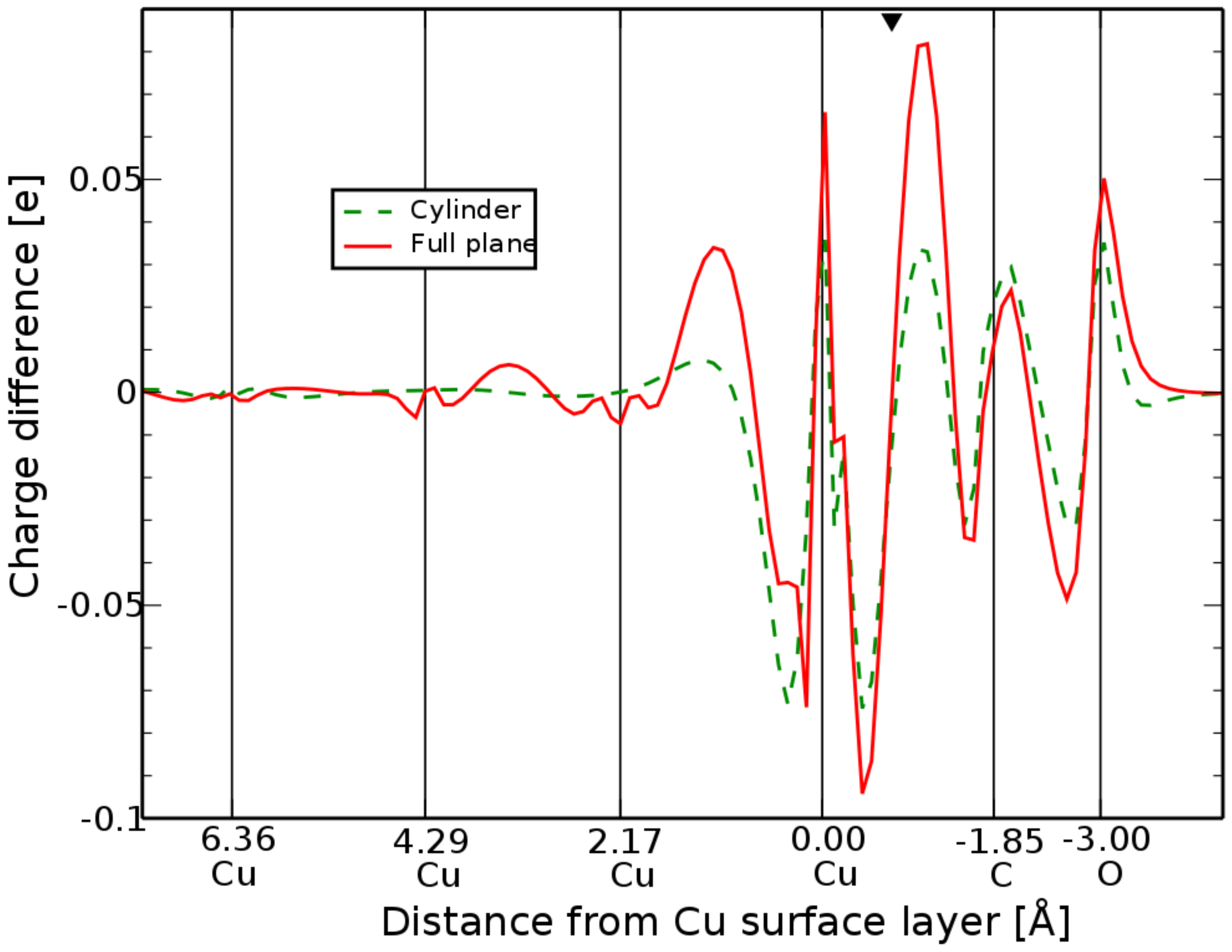}
\caption{Laterally integrated charge density difference curve of the $2\times2$ surface system. 
The red solid line denotes the integration over the full $xy$-plane and the green dashed line is restricted to 
a cylinder with 1\,Å diameter along a vertical axis that goes through the CO molecule. The triangle on top 
indicates the position of the Cu plane in Fig.~\ref{fig:elf}.
 \label{fig:chargediff}}
\end{figure}

To understand better how CO molecules modify the electronic structure of the Cu(111) surface, 
we have calculated the charge density difference between (i) the system with both a Cu surface (support) 
and a CO molecule (adsorbate) and (ii) and the corresponding separated systems:  
\begin{equation}
\rho_{\rm diff} = \rho_{\rm CO-Cu} -\rho_{\rm Cu}-\rho_{\rm CO}.
\end{equation}
The surface and the molecule are kept in exactly same positions as in the 
combined system while computing the separated charge densities. The laterally integrated 
charge density difference is computed with respect to the $xy$-plane, and it is displayed 
in the neighborhood of the Cu(111) surface for the $2\times2$ surface coverage in 
Fig.~\ref{fig:chargediff}. In addition, we have computed the charge density difference along the axis 
of the CO molecule by restricting the integration area within a cylinder of {0.5 \AA} radius.

The charge density difference in Fig.~\ref{fig:chargediff} shows oscillating features 
localized around the CO molecule and within the three uppermost Cu layers. The changes 
are largest at the interface. The chemical interaction between CO and Cu is visible as 
pronounced oscillations within the contact region (depletion close to Cu, accumulation close 
to C), but there is no significant charge transfer. The C=O bond has lost some charge while 
there is accumulation on both C and O atoms. The charge relocation effect is weak around 
the second and third Cu layer and vanishes deeper in the system. 

\begin{figure}
\includegraphics[scale=0.55]{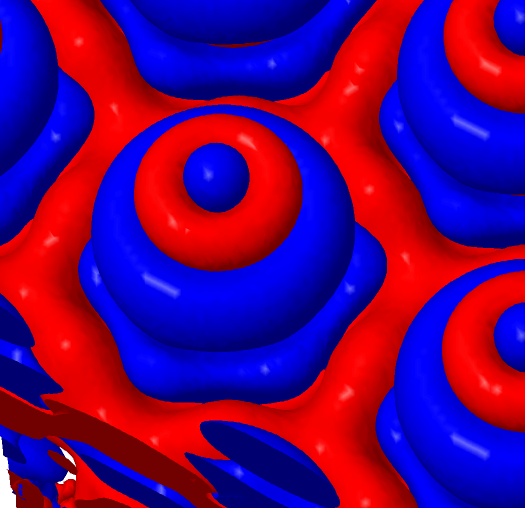}
\caption{Charge density difference isosurfaces (red, accumulation; blue, depletion) for the 
$2\times2$ coverage show how CO molecules push electron density to interstitial areas 
resulting in a hexagonal accumulation zone (red color).  
\label{fig:chdiff_surf}}
\end{figure}

Furthermore, we can investigate the 
lateral changes in the charge difference by comparing the integration over the full $xy$-plane 
and over a cylinder centered at the CO molecule axis. This analysis reveals a large charge 
accumulation zone {\it between} the CO molecules $\sim 0.9$\,Å above the top Cu layer, as 
the cylinder has less accumulation than the whole $xy$-plane. In other words, there is less 
charge accumulation right below the CO molecule (Cu-C contact) than in the surroundings 
due to a lateral charge-transfer effect. The corresponding electron accumulation zone 
between the CO molecules has a sixfold symmetry, and it is illustrated in Fig.~\ref{fig:chdiff_surf} 
as a three-dimensional isosurface visualization. The effect is not solely restricted above the
Cu surface, and a similar but smaller accumulation pattern can be seen between the Cu 
layers down to third Cu plane (see Fig.~\ref{fig:chargediff}). Therefore, we conclude that the
adsorption of CO results in a hexagonal charge accumulation pattern in the top layers of the 
Cu surface, which gives rise to graphene-like electronic properties. The lattice constant of 
this hexagonal network depends on the CO coverage. 

\begin{figure}
\includegraphics[scale=0.5]{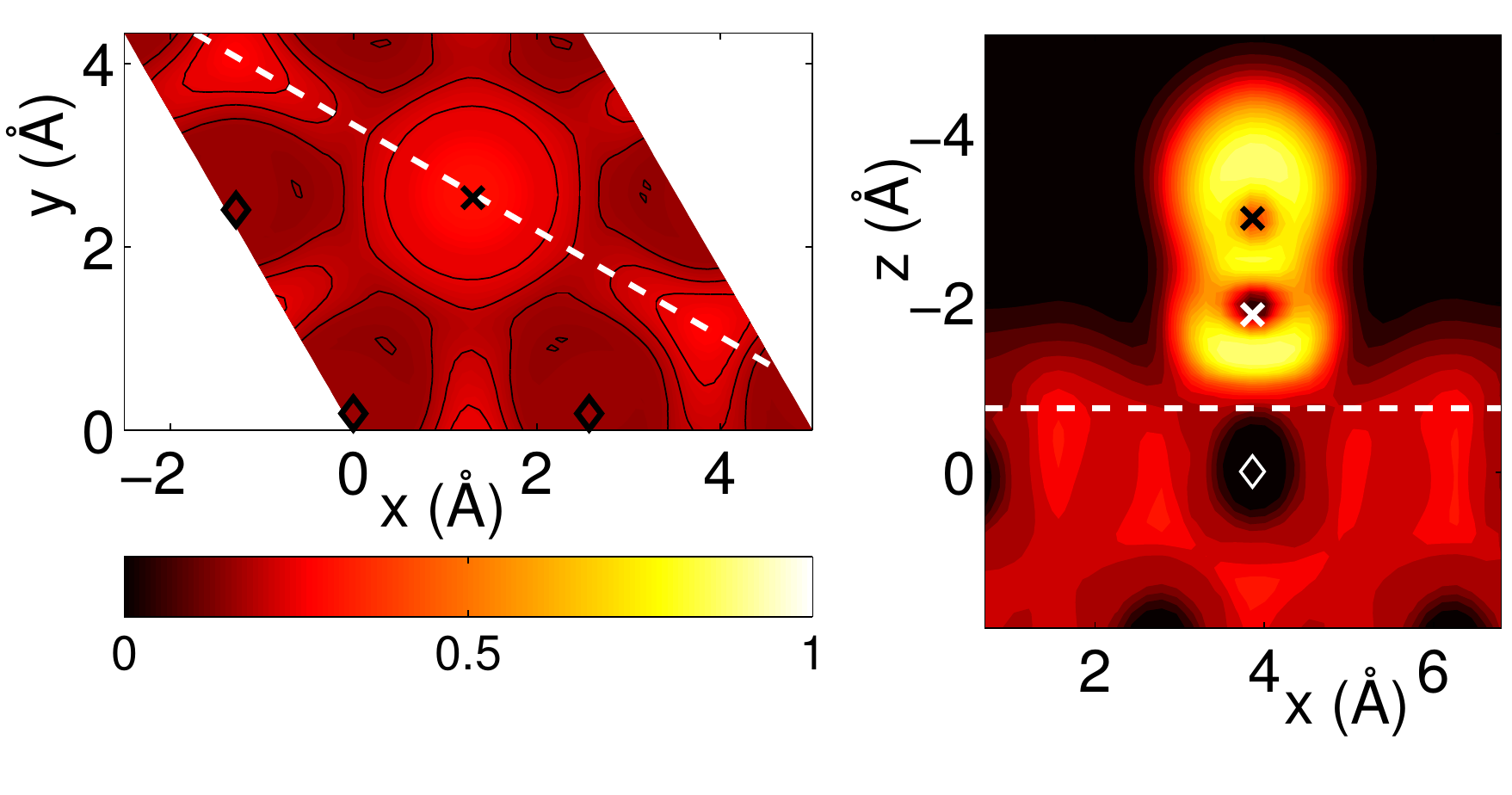}
\caption{Electron localization function (ELF) in colorscale for lateral ($xy$-plane, left) and vertical ($xz$-plane, 
right) cutplanes. The dotted white lines shows positions of the respective cutplanes in the neighboring panels. 
The crosses show the position of C and O atoms, and the diamonds show the Cu positions. Note that the system (unit cell) is periodically 
repeated.
\label{fig:elf}}
\end{figure}

In order to shed more light on the characteristics of the electron density in different locations, we 
plot the ELF~\cite{ss450_126} in Fig.~\ref{fig:elf} (colorscale) as cutplane presentations. The 
ELF gives a measure for electron pairing (localization) by establishing a renormalization of the Fermi hole
curvature. Its value varies between zero (no electron localization) and one 
(electron pairing, covalent bond). Due to employing PAWs in the calculations, the obtained 
ELF corresponds to the valence or pseudo-ELF,~\cite{jcc18_1431} which neglects the electron 
density of the atomic core. This explains why ELF is nearly zero at atomic sites. However, 
the pseudo-ELF should give results very similar to the all-electron ELF {\em outside} the atomic cores 
as the valence electrons are delocalized and responsible for chemical bonding. 

The covalent character of the C=O bond ELF is visible as a bright yellow color, and the C-Cu contact 
shows gradual reduction towards metallic bonding. This continuous feature, which is different from 
the ones observed for a simple charge transfer (ionic bonding) or van der Waals -type 
physisorption,~\cite{prb74_165404} confirms that the 
interaction between CO and Cu(111) should be described as chemisorption. The ELF analysis 
(Fig.~\ref{fig:elf}, right) also shows how CO pushes the ``metallic'' electron density at the surface 
to the intermediate zone between the adsorbate molecules. The lateral cut-plane presentation 
(Fig.~\ref{fig:elf}, left) shows further that the locations of these side lobes match with the 
corners of the hexagonal accumulation zones in Fig.~\ref{fig:chdiff_surf}, indicating that the
corresponding electron density has a delocalized (metallic) character.

\begin{figure}
\includegraphics[scale=0.2]{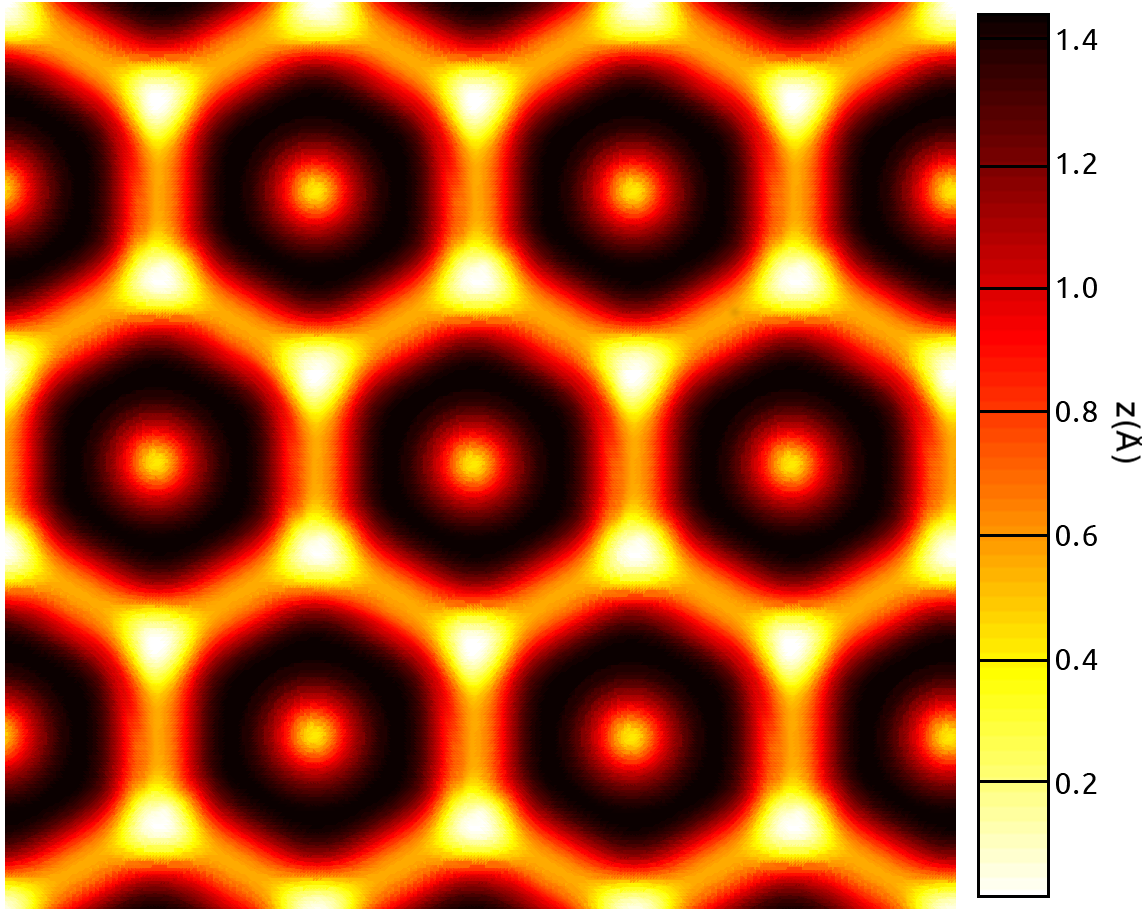}
\caption{Simulated STM image for the calculated with bias of 10\,mV for $2\times2$ surface 
using the Tersoff-Hamann scheme.~\cite{prb31_805} The figure illustrates how the flat areas 
between the CO molecules form a honeycomb lattice The zero of the scale is 4.0\,Å above the copper atom under the CO molecule.
 \label{fig:stm4}}
\end{figure}

Finally, we have simulated the STM images using the Tersoff-Hamann scheme \cite{prb31_805} for 
the $2\times2$ surface with 10\,mV bias voltage. The result is shown in Fig.~\ref{fig:stm4}. 
The STM image illustrates the hexagonal structure of the surface, i.e., the electrons
are expected to primarily move between the CO molecules that form scattering centers
similar to those previously modeled in 2D in Ref.~\onlinecite{flake}. The image is very similar to the
experimental STM images by Gomes \emph{et al.}\cite{nature483_306} However, we need to 
bear in mind that the height difference between CO molecules and intemediate areas is reversed. 

To summarize, we have used DFT simulations to investigate
a particular form of artificial graphene, carbon monoxide molecules on a copper (111) 
surface. We observe a Dirac point in the electronic band structure near the Fermi energy 
at the K-point for different CO coverages.
A detailed analysis of the charge density and electron localization show us that the CO 
adsorption (chemisorption) introduces a lateral charge accumulation in the top layers of 
Cu(111) with a hexagonal pattern. The associated electron density has a metallic 
character, and the honeycomb lattice constant is governed by the CO coverage. The 
theoretical Fermi velocities and STM images are in a reasonable agreement with the
experiments, confirming further the validity of the DFT approach. Our first-principles 
approach can be readily used to study other variants of molecular graphene in the quest
of designer Dirac materials.

\section{Acknowledgements}
This work has been supported by the Academy of Finland through its Centre of Excellence 
Program (Project no. 251748) and through Project No. 126205, 
European Community's FP7 through the CRONOS project, Grant Agreement No. 280879, and  
COST Action CM1204 (XLIC). The computer resources of the Finnish IT Center for Science (CSC) and 
Finnish Grid Infrastructure (FGI) are acknowledged. 

\section{References}

\end{document}